\documentclass[useAMS,usenatbib]{mn2e}
\usepackage{graphicx}
\usepackage{pdflscape}
\usepackage{breqn}
\usepackage{subfig}
\title[Chemical abundances in a sample of solar analogues]{Mg, Al, Si, Ca, Ti, Fe, and Ni abundance for a sample of solar analogues}

\author[L{\'o}pez-Valdivia R., Bertone E. \& Ch\'avez M. ]{Ricardo L{\'o}pez-Valdivia\thanks{E-mail:
valdivia@inaoep.mx},  Emanuele Bertone, and Miguel Ch{\'a}vez \\
Instituto Nacional de Astrof{\'i}sica, {\'O}ptica y Electr{\'o}nica, Luis Enrique Erro 1, Tonantzintla, Puebla, 72840, M{\'e}xico}

\begin{document}

\date{Accepted XXX. Received YYY; in original form ZZZ}

\pagerange{\pageref{firstpage}--\pageref{lastpage}} \pubyear{2016}

\maketitle

\label{firstpage}

\begin{abstract}
We report on the determination of chemical abundances of 38 solar analogues, including 11 objects previously identified as super metal-rich stars. We have measured the equivalent widths for 34 lines of 7 different chemical elements 
(Mg, Al, Si, Ca, Ti, Fe, and Ni) in high-resolution ($\mathcal{R} \sim 80\,000$) spectroscopic images, obtained 
at the Observatorio  Astrof\'isico Guillermo Haro (Sonora, Mexico), with the Cananea High-resolution 
Spectrograph. We derived chemical abundances using  ATLAS12 model 
atmospheres and the Fortran code MOOG. 
We confirmed the super metallicity status of 6 solar analogues. Within our sample, BD+60~600 is the most metal-rich star ([Fe/H]=+0.35~dex), while for HD~166991 we obtained the lowest iron abundance 
([Fe/H]=$-0.53$~dex).  We also computed the so-called [Ref] index for 25 of our solar analogues, and we 
found, that BD+60~600 ([Ref]=+0.42) and BD+28~3198 ([Ref]=+0.34) are good targets for exoplanet search.
\end{abstract}


%***********************************
%INTRODUCTION
%************************************
\section{Introduction}
Stellar chemical composition represents an important parameter in stellar  and galactic
astronomy studies, and, in particular, in the relatively recent field of exoplanets.
In this latter field, different studies have aimed at searching for possible correlations between  
properties (mainly chemical composition) of host stars and the occurrence of exoplanets. \cite{gonzalez97}, 
with the search for exoplanets still in its early stages, suggested a link between high metal content of host 
stars and the presence of giant gaseous planets. Such correlation was later confirmed by other authors (e.g., 
\citealp{fischer05}; \citealp{johnson10}; \citealp{sousa11}) and it agrees with the core accretion theory 
for planet formation (\citealp{pollack96}; \citealp{alibert04}), where high metallicity facilitates the 
formation of giant gas planets.

Within this scenario, the iron abundance ([Fe/H]) is commonly used as proxy for overall 
metallicity; however, \cite{gonzalez09} suggested the use of a new metallicity index, called [Ref], which 
takes into account the mass abundance of the refractory elements Mg, Si, and Fe, since their number densities 
and condensation temperatures are very similar. This [Ref] index is more sensitive (mainly at values greater 
than +0.20~dex) than [Fe/H] to describe the incidence probability of giant planets orbiting a star 
\citep{gonzalez14}.

The present work is the continuation of a global project aimed at determining atmospheric parameters and 
chemical abundances of solar analogues (main sequence stars with spectral types between G0 and 
G3)\footnote{In our sample the stars HD~130948 and HD~168874 have a different spectral type; nevertheless, we 
included them, because their atmospheric parameters are compatible with the rest of the sample.}, with 
special interest in looking for giant exoplanet host star candidates. In \cite{lopezval14}, we 
simultaneously determined the basic stellar atmospheric parameters [effective temperature ($T_{\rm eff}$), surface 
gravity ($\log g$), and global metallicity ([M/H])], for a sample of 233 solar analogues, using intermediate-resolution spectra ($\mathcal{R} \sim 1700$ at 4300~\AA) and a set of Lick-like indices defined within 
3800--4800~\AA. We determined for the first time the atmospheric parameters for 213 stars, of which 20 
are new super metal-rich star candidates (SMR; \rm{[M/H]}$\geq$0.16~dex).

The second goal of our project is the analysis of chemical abundances, which we started with 
the determination of the lithium abundance of a sample of 52 stars \citep{rlopezval15}. The analysis was 
carried out using narrow band high-resolution spectra ($\mathcal{R} \sim 80\,000$) centred on the 6708~\AA\ lithium feature. This sample included 12 SMR objects from our previous work \citep{lopezval14}.

In this third part of the series, we complement the lithium abundance with the chemical abundances of Mg, Al, 
Si, Ca, Ti, Fe, and Ni, for 38 solar analogues. The sample and the observations are described in Section 2. 
In Section 3, we detail the determination of the chemical abundances, and, in Section 4, we discuss the 
results.
%%%%%%%%%%%%%%%%%%%%%%%%%%%%%%%%%%%%%%%%%%%%%%%%%
%sample
%%%%%%%%%%%%%%%%%%%%%%%%%%%%%%%%%%%%%%%%%%%%%%%%% 
\section{Stellar sample and Observations}
We selected 38 objects among the brightest stars of \citet{rlopezval15}. In Table~\ref{tab:samp}, we list the
name of the star, the visual magnitude, the spectral type and the atmospheric parameters (and their 
uncertainties) for the entire sample. The spectroscopic data were collected at the 2.1~m telescope of the 
Observatorio Astrof\'isico Guillermo Haro, located in Mexico, using the Cananea High-resolution Spectrograph 
(CanHiS). CanHiS is equipped with mid-band filters, that provide access to  $\sim$40~\AA\ wide wavelength 
intervals in a single diffraction order.

\begin{figure*}
\centering
\includegraphics[width = 160mm]{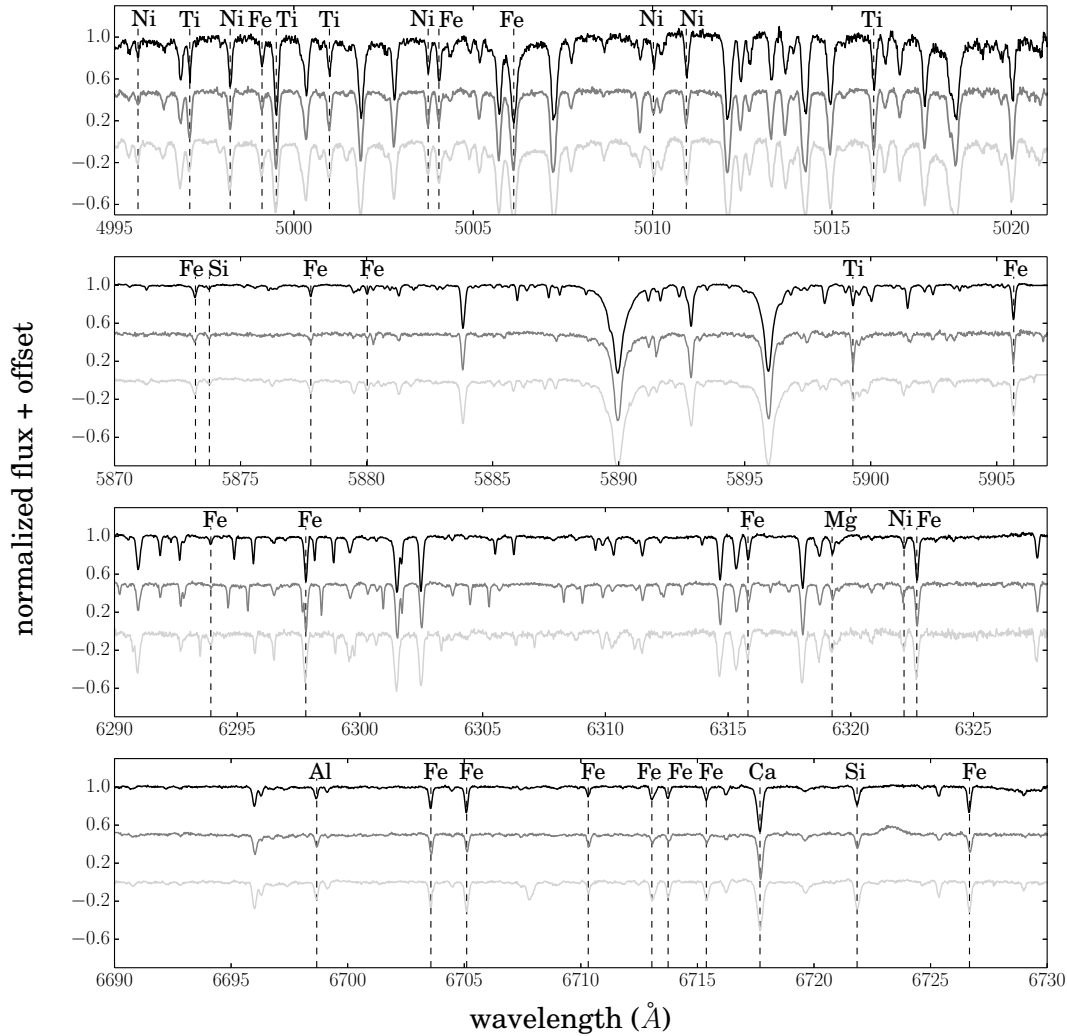}
\caption{Normalized spectra of Vesta (black), HD 12699 (gray), and BD+28 3198 (light gray) in the four spectral regions, with identification of the atomic lines used to compute abundances. Note that the interval centred at 6310~\AA\ is more affected by the presence of telluric lines.}
\label{fig:lin}
\end{figure*}

We observed the entire sample with a spectral resolving power of $\mathcal{R}\sim 80\,000$ and a typical 
signal-to-noise ratio (S/N) of about 100, using 4 different filters of CanHiS, centred at 5005, 
5890, 6310, and 6710~\AA, respectively, giving  access to lines of Mg, Al, Si, Ca, Ti, Fe, and Ni 
(Fig.~\ref{fig:lin}). We also obtained the solar spectrum reflected by the asteroid Vesta with the same
instrumental setup. Per filter and per star, we collected at least 3 exposures, resulting in total exposure 
times between 1.5 and 3 hours.

Data reduction was conducted following the standard procedures of IRAF: bias subtraction, flat-field 
correction, cosmic-ray removal, wavelength calibration through an internal UNe lamp, and, finally, continuum 
normalization. We then shifted all the spectra to the rest frame, using a degraded (to our resolution) 
version of the high-resolution spectrum  of the Sun \citep{kurucz84} as template. For each star (and filter) 
we co-added single exposures weighted by the S/N to obtain the final spectrum.

\begin{table*}
\begin{minipage}{155mm}
\caption{Atmospheric parameters and their uncertainties of the stellar sample. For all the stars of our sample the error on the microturbulence velocity is 0.27~km\,s$^{-1}$.}
\label{tab:samp}
\begin{tabular}{lrlrrrrrrc}
\hline
Object  &   V   &  SType & $T_{\rm eff}$  & $\sigma$ & $\log g$ & $\sigma$  & $\rm{[M/H]}$ & $\sigma$ & $\xi$ \\
&   &   &  (K)   & (K) & (dex) & (dex) & (dex) & (dex) & (km\,s$^{-1}$) \\
\hline  
HD 5649	         &	8.70	&	G0V	&	5830	&	52	&	4.45	&   0.22  &  -0.08  &   0.04  & 1.03  \\
BD+60 402        &     10.26    &	G0V	&	5985	&	72	&	4.30	&   0.40  &   0.22  &   0.09  & 1.15  \\
HD 16894         &	8.02	&	G2V	&	5500	&	70	&	4.05	&   0.30  &  -0.10  &   0.09  & 0.83  \\
BD+60 600        &	8.65	&	G0V	&	5655	&	47	&	3.95	&   0.20  &   0.20  &   0.07  & 1.18  \\
HD 232824        &	9.52	&	G2V	&	5900	&	67	&	4.15	&   0.35  &   0.16  &   0.08  & 1.27  \\
HD 237200        &	9.66	&	G0V	&	6045	&	55	&	4.25	&   0.32  &   0.18  &   0.05  & 1.26  \\
HD 26710         &	7.18	&	G2V	&	5815	&	47	&	4.55	&   0.20  &  -0.04  &   0.04  & 1.63  \\
HD 31867         &	8.05	&	G2V	&	5590	&	57	&	4.40	&   0.25  &  -0.10  &   0.06  & 0.98  \\
HD 33866         &	7.87	&	G2V	&	5481	&	123	&	4.33	&   0.27  &  -0.07  &   0.10  & 0.91  \\
HD 41708         &	8.03	&	G0V	&	5998	&	58	&	4.55	&   0.27  &   0.08  &   0.10  & 1.33  \\
HD 42802         &	6.44	&	G2V	&	5617    &	80	&	4.53	&   0.27  &  -0.11  &   0.10  & 0.98  \\
HD 77730         &	7.39	&	G2V	&	5698    &	80	&	4.13	&   0.27  &  -0.05  &   0.10  & 1.00  \\
HD 110882        &	8.87	&	G1V	&	5880	&	50	&	4.40	&   0.25  &  -0.28  &   0.04  & 1.14  \\
HD 110884        &	9.11	&	G3V	&	5905	&	87	&	4.30	&   0.40  &  -0.26  &   0.08  & 1.16  \\
HD 111513        &	7.35	&	G1V	&	5723    &	80	&	4.31	&   0.27  &   0.12  &   0.10  & 1.21  \\
HD 111540        &	9.54	&	G1V	&	5840	&	47	&	4.20	&   0.25  &   0.14  &   0.05  & 1.13  \\
HD 124019        &	8.56	&	G2V	&	5685	&	57	&	4.65	&   0.25  &  -0.18  &   0.06  & 0.88  \\
HD 126991        &	7.90	&	G2V	&	5360	&	107	&	3.15	&   0.40  &  -0.34  &   0.14  & 1.27  \\
HD 129357        &	7.83	&	G2V	&	5775	&	52	&	4.30	&   0.22  &  -0.14  &   0.05  & 1.21  \\
HD 130948        &	5.88	&   F9IV-V      & 	5885    &	80	&	4.42	&   0.27  &  -0.09  &   0.10  & 1.28 \\
HD 135145        &	8.35	&	G0V	&	5997	&	80	&	4.14	&   0.27  &  -0.02  &   0.10  & 1.19  \\
HD 135633        &	8.46	&	G0V	&	6095	&	67	&	4.25	&   0.40  &   0.22  &   0.06  & 1.27  \\
HD 140385        &	8.57	&	G2V	&	5735	&	60	&	4.60	&   0.27  &  -0.16  &   0.08  & 1.13  \\
HD 145404        &	8.54	&	G0V	&	5920	&	82	&	4.43	&   0.27  &  -0.16  &   0.10  & 1.20  \\
HD 152264        &	7.74	&	G0V	&	6177	&	73	&	4.09	&   0.27  &   0.02  &   0.10  & 1.36  \\
BD+29 2963       &	8.42	&	G0V	&	5865	&	55	&	4.70	&   0.22  &   0.00  &   0.04  & 1.27  \\
HD 156968        &	7.97	&	G0V	&	6105	&	96	&	4.42	&   0.27  &  -0.03  &   0.10  & 1.30  \\
HD 168874        &	7.01	&	G2IV	&	5696    &	80	&	4.41	&   0.27  &  -0.05  &   0.10  & 0.86  \\
BD+28 3198       &	8.66	&	G2V	&	5840	&	35	&	4.00	&   0.17  &   0.24  &   0.05  & 1.25  \\
TYC 2655-3677-1	 &  9.93&	G0V &	6220	&	47	&       4.15	&   0.27  &   0.28  &   0.05  & 1.31  \\
HD 333565        &	8.75	&	G1V	&	5990	&	52	&	4.45	&   0.27  &   0.12  &   0.05  & 1.19  \\
HD 228356        &	9.07	&	G0V	&	6055	&	37	&	4.00    &   0.20  &   0.16  &   0.05  & 1.41  \\
HD 193664        &	5.93	&	G3V	&	5942    &	112	&	4.47	&   0.27  &  -0.11  &   0.10  & 1.28  \\
BD+47 3218       &	8.70	&	G0V	&	6050	&	52	&	4.05	&   0.30  &   0.16  &   0.06  & 1.41  \\
HD 210460        & 6.19 &   G0V  &   5357&   80  &       3.58    &   0.27  &  -0.17  &   0.10  & 1.27  \\
TYC 3986-3381-1	 & 10.37&	G2V	&	5855	&	57	&	4.15	&   0.25  &   0.26  &   0.07  & 1.15  \\
HD 212809	 &	8.64	&	G2V	&	5975	&	55	&	4.55	&   0.27  &   0.16  &   0.05  & 1.33  \\
BD+28 4515	 &	8.73	&	G2V	&	5580	&	40	&	3.50	&   0.17  &  -0.22  &   0.06  & 1.29  \\
\hline
\end{tabular}
\end{minipage}
\end{table*}
%%%%%%%%%%%%%%%%%%%%%%%%%%%%%%%%%%%%%%%%%
%abundances
%%%%%%%%%%%%%%%%%%%%%%%%%%%%%%%%%%%%%%%%%
\section{Abundances Determination}
\label{sect:abun}
We determined the chemical abundances, through a local thermodynamic equilibrium (LTE) analysis, using  the driver {\it abfind} of the February 
2013 version of MOOG \citep{sneden73}, which performs an adjustment of the abundance to 
match a single-line equivalent width (EW). MOOG requires a standard solar composition (we used the solar 
abundances of \citealt{grevesse98}), a model atmosphere, a line list, and an EW measurement to compute 
atomic abundances. Below we describe in detail each of these requirements.

\subsection{Photospheric parameters and model atmospheres}
\label{model}
In order to compute a model atmosphere the basic parameters are required: $T_{\rm eff}$, 
$\log{g}$, [M/H], and the microturbulence velocity ($\xi$). We adopted the $T_{\rm eff}$, 
$\log{g}$, and [M/H] values of our previous work \citep{rlopezval15}. For $\xi$, we used the grid of 
atmospheric parameters of \cite{takeda05}, which includes determination of $T_{\rm eff}$, $\log{g}$, [M/H], 
and $\xi$ for 160~FGK stars. We looked within the Takeda's grid the nearest set of the first 3  parameters 
for each star in our sample, and we assigned the Takeda's determination of $\xi$ to our star. We found $\xi$ 
values between 0.83 and 1.63~km\,s$^{-1}$, which are in agreement with values determined from synthetic 
spectra \citep{husser13}.

Regarding the atmospheric parameters uncertainties, we used those reported in \cite{rlopezval15}. For those 
cases where uncertainties were not available, we assigned, for $\log g$ and $\xi$, $\pm$~0.27~dex and 
$\pm$~0.27~km\,s$^{-1}$, as the typical uncertainty, which is the standard deviation of both $\log g$ and 
$\xi$ distributions of the Takeda's stars with $T_{\rm eff}$ within the values of our sample. For the 
uncertainty of [M/H] we assumed $\pm$0.10~dex as a conservative error. 

Using the atmospheric parameters reported in Table~\ref{tab:samp}, we computed an ATLAS12 \citep{kurucz13} 
model atmosphere for each star; we also computed a solar model atmosphere with 
$T_{\rm eff,\odot}$=5777~K, $\log{g}_\odot$=4.44~dex, [M/H]$_\odot$=0.0~dex, and $\xi_\odot$=1.0~km\,s$^{-1}$. 

%%%%%%%%%%%%%%%%%%%%%%%%%%%%
\subsection{Line list}
We extracted the atomic transitions between 4995 and 6730~\AA\ from The Viena Atomic Line Database (VALD, 
\citealp{piskunov95}; \citealp{kupka99}), using the atmospheric parameters of the Sun. 
With these atomic transitions and the ATLAS12 solar model, we created with SYNTHE (\citealp{kurucz79}; 
\citealp{kurucz81}; \citealp{kurucz93}) a synthetic solar spectrum at the same spectral resolution as our 
observations. From the Vesta spectrum we selected 34 suitable atomic lines (listed in Table~\ref{tab:ato} and 
shown in Fig.~\ref{fig:lin}) of 7 different chemical elements (Mg, Al, Si, Ca, Ti, Fe, and Ni), avoiding 
weak or saturated lines and blends.
 
\cite{neves09} pointed out that oscillator strengths ($\log gf$) of VALD might not be accurate enough 
for all the atomic transitions. To correct these possible inaccuracies, we determined the EW (see 
Section~\ref{sect:ew}) for the 34 selected lines in the observed and synthetic solar spectrum; then, we 
compared both measurements and we modified the $\log gf$ until both measurements (observed and synthetic) 
agreed. For 15 lines, we also slightly modified the central wavelength reported by VALD. The transition 
parameters from VALD as well as their modifications are reported in Table~\ref{tab:ato}.

\begin{table}
\caption{Set of atomic parameters from VALD and the modifications made by us to the central wavelength and the $\log{gf}$.}
\label{tab:ato}
\begin{tabular}{|cccccc|}\hline
$\lambda$ & $\Delta$ $\lambda$ & element& $\chi$ &  $\log gf$  & $\Delta$ $\log gf$ \\ 
  (\AA)   &  (\AA) &        &  (eV)         &   &   \\ \hline
4995.650   &  0.005    &	Ni~\scriptsize{I}	&	3.635	&	-1.580	&   -0.308  \\
4997.098   &    -      &	Ti~\scriptsize{I}	&	0.000	&	-2.070	&   -0.156  \\
4998.224   &    -      &	Ni~\scriptsize{I}	&	3.606	&	-0.700	&   -0.261  \\
4999.112   &    -      &	Fe~\scriptsize{I}	&	4.186	&	-1.740	&   -0.066  \\
4999.503   &  0.007    &	Ti~\scriptsize{I}	&	0.826	&	0.320	&   -0.279  \\
5000.990   &  0.002    &	Ti~\scriptsize{I}	&	1.997	&	-0.020	&   -0.255  \\
5003.741   &  0.003    &	Ni~\scriptsize{I}	&	1.676	&	-3.070	&   -0.265  \\
5004.044   &    -      &	Fe~\scriptsize{I}	&	4.209	&	-1.400	&   -0.110  \\
5006.119   &  0.011    &	Fe~\scriptsize{I}	&	2.833	&	-0.638	&   -0.336  \\
5010.023   &    -      &	Ni~\scriptsize{I}	&	3.768	&	-0.980	&   -0.085  \\
5010.938   &  0.002    &	Ni~\scriptsize{I}	&	3.635	&	-0.870	&   -0.161  \\
5016.161   &  0.004    &	Ti~\scriptsize{I}	&	0.848	&	-0.480	&   -0.294  \\
5873.212   &    -      &	Fe~\scriptsize{I}	&	4.256	&	-2.140	&    0.168  \\
5873.763   &    -      &	Si~\scriptsize{I}	&	4.930	&	-4.244	&    1.194  \\
5877.788   &    -      &	Fe~\scriptsize{I}	&	4.178	&	-2.230	&   -0.009  \\
5880.027   &    -      &	Fe~\scriptsize{I}	&	4.559	&	-1.940	&   -0.028  \\
5899.293   &    -      &	Ti~\scriptsize{I}	&	1.053	&	-1.100	&   -0.098  \\
5905.671   &  0.003    &	Fe~\scriptsize{I}	&	4.652	&	-0.730	&   -0.179  \\
6293.925   &    -      &	Fe~\scriptsize{I}	&	4.835	&	-1.717	&   -0.083  \\
6297.792   &  0.002    &	Fe~\scriptsize{I}	&	2.223	&	-2.740	&   -0.185  \\
6315.811   &    -      &	Fe~\scriptsize{I}	&	4.076	&	-1.710	&   -0.023  \\
6319.237   &    -      &	Mg~\scriptsize{I}	&	5.108	&	-2.324	&    0.238  \\	
6322.166   &  0.003    &	Ni~\scriptsize{I}	&	4.154	&	-2.426	&    1.267  \\
6322.685   &  0.004    &	Fe~\scriptsize{I}	&	2.588	&	-1.170	&   -1.256  \\
6698.673   &    -      &	Al~\scriptsize{I}	&	3.143	&	-1.647	&   -0.255  \\
6703.566   &  0.003    &	Fe~\scriptsize{I}	&	2.759	&	-3.160	&    0.097  \\
6705.101   &  0.003    &	Fe~\scriptsize{I}	&	4.607	&	-1.392	&    0.269  \\
6710.318   &    -      &	Fe~\scriptsize{I}	&	1.485	&	-4.880	&    0.036  \\
6713.046   &    -      &	Fe~\scriptsize{I}	&	4.607	&	-0.963	&   -0.380  \\
6713.743   &    -      &	Fe~\scriptsize{I}	&	4.796	&	-1.600	&    0.186  \\
6715.382   &    -      &	Fe~\scriptsize{I}	&	4.608	&	-1.640	&    0.109  \\
6717.681   &  0.003    &	Ca~\scriptsize{I}	&	2.709	&	-0.524	&    0.025  \\
6721.848   &    -      &	Si~\scriptsize{I}	&	5.863	&	-1.527	&    0.415  \\
6726.666   &  0.003    &	Fe~\scriptsize{I}	&	4.607	&	-1.133	&    0.078  \\
\hline
\end{tabular}
\end{table}

\subsection{Equivalent widths}
\label{sect:ew}
The EW determination plays a fundamental role in the abundance 
determination. Since the EW depends strongly on the local continuum level, it is of crucial importance to
determine it as accurately as possible. We implemented the following procedure to establish the local 
continuum level and to measure the EW.

First, by means of a Gaussian fit of a small region (5~\AA), we identified and removed the points that 
form the spectral line of interest, which are points enclosed in a interval of $\pm3\sigma$ from the central 
wavelength of the line. Then, we passed through an iterative routine the remaining spectrum, which is 
a combination of neighbouring lines and noise, to remove points above $\pm$2$\sigma$ their average value in order to identify the local continuum. Finally, we adjusted to the line a Gaussian profile whose integral represents its EW.

We estimated the error on the EW applying a Monte Carlo method with 1000 iterations, randomly adding to
the spectrum the noise of the local continuum.

\begin{figure}
\centering
\includegraphics[width = 80mm]{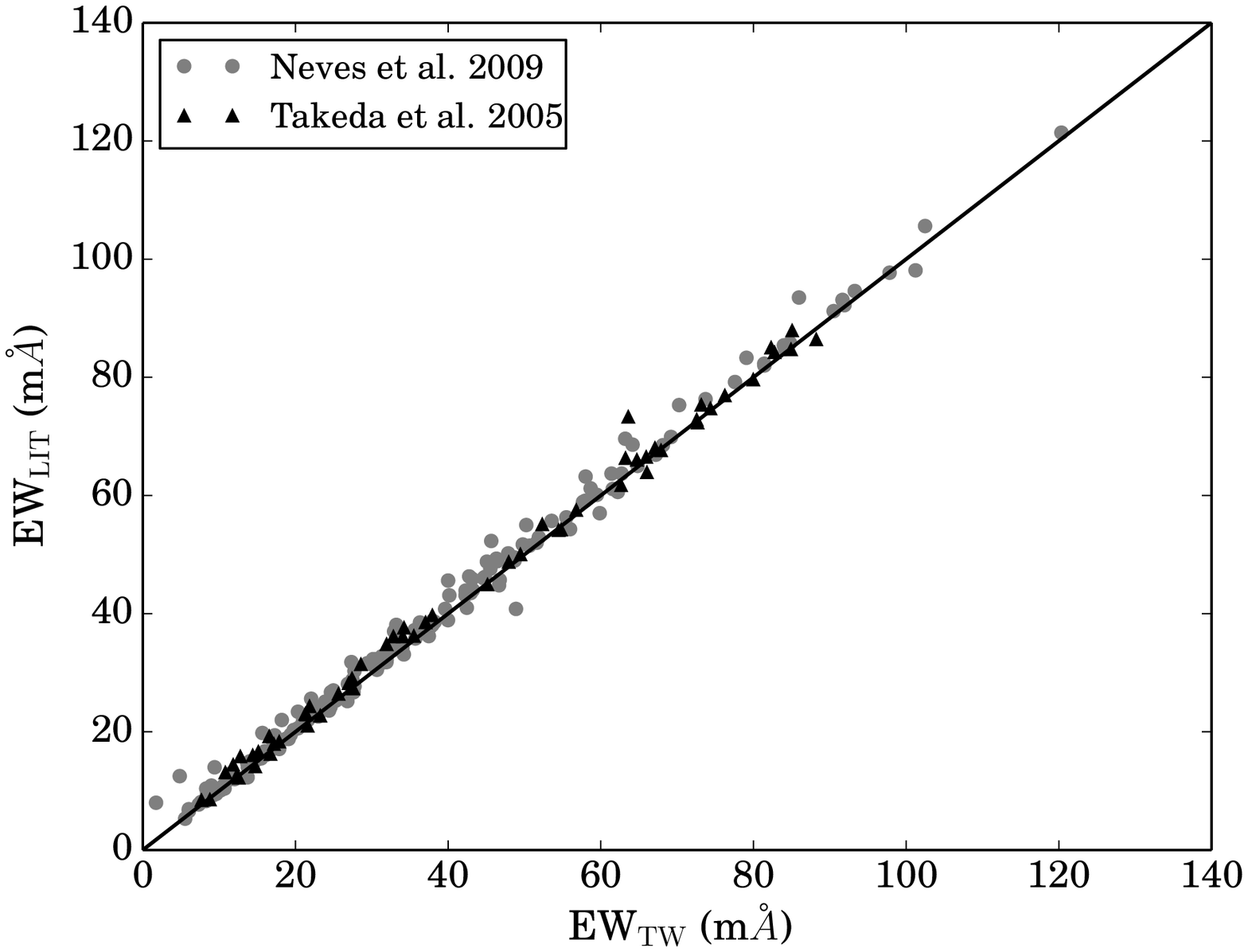}
\caption{Comparison of the solar EW computed in this study (EW$_{\mathrm{TW}}$) and those determined by \citet{takeda05} and \citet{neves09}.}
\label{fig:c_ew}
\end{figure}

We checked the consistency of our procedure by means of a comparison of solar line EWs determined in two 
different works (\citealt{neves09}; \citealt{takeda05}) with those determined by us. 
We measured in the solar spectrum of \cite{kurucz84}, also used by Takeda and Neves, the EW for 57 and 
178 iron lines reported by \citealt{takeda05} and \citealt{neves09}, respectively. From this comparison, 
which is depicted in Fig.~\ref{fig:c_ew}, we found good agreement, with some small differences, 
which can be explained by different local continuum levels.

\subsection{Abundances computation and error budget.}

For each star and Vesta, we measured the EW of all lines listed in Table~\ref{tab:ato}.  
We rejected, through visual inspection, the lines whose best fit was not accurate enough; these lines
vary from star to star.
The EWs of Table~\ref{tab:ew} were used in MOOG to compute the chemical abundances.  
For species with more than one analysed transition, we carried out a weighted mean to 
obtain the final abundance, after having discarded outliers with an iterative 3$\sigma$ clipping.

It is important to note that these two rejection processes could introduce potential biases and 
different abundance scales in stars with different excluded lines. The first filter is actually a visual 
inspection that relies on the S/N of the spectra and is not directly associated with abundances, while the 
sigma clipping is indeed applied directly to abundances, but it was employed in only one Fe line of eight 
stars. In order to take into account these potential biases, we conducted a Monte Carlo procedure in
which we computed the Fe, Ni, and Ti abundance (elements with more available lines within our line list with 
18, 6, and 5, respectively) for Vesta and some stars of our sample. We computed the mean abundance of Fe, Ni, 
and Ti using different size sets of randomly selected lines. After 1000 iterations for each set, element, and 
star, we demonstrated that the final abundance of these elements in all the cases does not change by more 
than 0.02 dex on average.\\

We report in Table~\ref{tab:abun} the  abundances of the 7 atomic elements for our sample; they are given 
with respect to the solar abundances determined for 
Vesta\footnote{[X/H] = A(X)$_{\rm star}$ - A(X)$_{\odot}$, where A(X)$_{\odot}$ is the computed abundance 
for Vesta.} (see Table~\ref{tab:ref}). The Table also provides the abundance uncertainty and the number of 
lines used for the abundance determination. \\\\

Along with the uncertainty on 
the EW measurement, the error on the stellar parameters is the source that most affects the final abundances. 
To properly assess it, we constructed a small matrix of abundance variations as a function of the difference 
in four atmospheric parameters ($T_{\rm eff}$, $\log{g}$, [M/H], and $\xi$), taking the solar values as 
reference. For each absorption line $j$,  we considered the EW measured on the Vesta spectrum and we computed 
a grid of abundance variations $\Delta{\rm[X/H]_{j}} = {\rm[X/H]_{j}} - {\rm[X/H]_{j,\odot}}$, caused by a 
difference $\Delta T_{\rm eff,\odot}=150$~K, of $\Delta\log{g}_\odot=0.40$~dex, of $\Delta{\rm 
[M/H]}_\odot=0.20$~dex, and of $\Delta\xi_\odot=0.50$~km\,s$^{-1}$. Then, for each star, we obtained the 
$\Delta{\rm[X/H]_{j}}$ corresponding to each atmospheric parameter by linearly interpolating this grid, 
assuming, as parameter value difference, the errors reported in Table~\ref{tab:samp}. The error on the 
abundance derived from each absorption line is the quadratic sum of the error on the atmospheric parameters 
and the EW.\newpage

\begin{table}
\caption{EWs of the atomic lines considered for abundance determination. The complete Table is available in electronic version.}
\label{tab:ew}
\begin{tabular}{cccrc}
\hline
Star & $\lambda$ & Ele. & EW & $\sigma$ \\
	 & (\AA) & & (m\AA) & (m\AA) \\
\hline
         HD 5649 &  4998.224 &  28 &  27.35 &  2.51  \\
         HD 5649 &  4999.510 &  22 &  59.01 &  2.77  \\
         HD 5649 &  5000.992 &  22 &  24.08 &  3.18  \\
         HD 5649 &  5006.130 &  26 &  124.59 &  5.58 \\
         HD 5649 &  5010.940 &  28 &  23.02 &  2.81  \\
         HD 5649 &  5016.165 &  22 &  24.09 &  3.35  \\
         HD 5649 &  5877.788 &  26 &  8.71 &  1.16   \\
         HD 5649 &  5880.027 &  26 &  10.47 &  1.57  \\
         HD 5649 &  5905.674 &  26 &  30.81 &  1.62  \\
         HD 5649 &  6297.794 &  26 &  47.92 &  1.26  \\
         HD 5649 &  6322.689 &  26 &  49.03 &  1.82  \\
         ... & ...  & ... & ...  &  ... \\
\hline
\end{tabular}
\end{table}

\section{Discussion}

\begin{table*}
\caption{Chemical abundances of the stellar sample. For each element we present in different rows 
the weighted mean abundance, its error, and the number of lines used in the determination of the abundance 
per star.}
\begin{tabular}{lrrrrrrr} \hline
Star & [Mg/H] & [Al/H] & [Si/H] & [Ca/H] & [Ti/H] & [Fe/H] & [Ni/H]\\ \hline  
            HD 5649 & -- & -- & -- & -0.48 &-0.42  & -0.32 & -0.43 \\
                   &  -- &  -- &  -- &  0.13 & 0.07  &  0.02 &  0.06 \\
                   &  --    &  --    &  --    &  1    & 3     &  9    &  2    \\
          BD+60 402 & +0.18 & +0.20 & +0.20 & +0.26 &+0.17  & +0.19 & +0.16 \\
                   & 0.12  & 0.04  & 0.04  & 0.19  &0.06   & 0.02  & 0.04  \\
                   &  1    &  1    &  1    &  1    & 4     & 15    &  5    \\
           HD 16894 & -0.21 & -- & +0.25 & +0.32 &+0.07  & +0.04 & -0.01 \\
                   &  0.05 &  -- &  0.08 &  0.17 & 0.06  &  0.02 &  0.04 \\
                   &  1    &  --    &  1    &  1    & 4     & 16    &  6    \\
          BD+60 600 & +0.51 & +0.39 & +0.44 & +0.51 &+0.24  & +0.35 & +0.34 \\
                   &  0.05 &  0.04 &  0.04 &  0.13 & 0.05  &  0.01 &  0.03 \\
                   &  1    &  1    &  2    &  1    & 4     & 18    &  5    \\
          HD 232824 & -0.07 & -0.08 & -0.03 & -0.07 &-0.24  & -0.09 & -0.06 \\
                   &  0.07 &  0.05 &  0.04 &  0.17 & 0.07  &  0.02 &  0.04 \\
                   &  1    &  1    &  2    &  1    & 4     & 17    &  4    \\
          HD 237200 & +0.03 & +0.13 & +0.20 & +0.27 &+0.21  & +0.18 & +0.14 \\
                   &  0.07 &  0.04 &  0.04 &  0.17 & 0.07  &  0.01 &  0.04 \\
                   &  1    &  1    &  2    &  1    & 4     & 16    &  5    \\
           HD 26710 & -0.13 & -- & +0.09 & +0.08 &-0.23  & +0.08 & -0.12 \\
                   &  0.04 &  -- &  0.04 &  0.13 & 0.05  &  0.01 &  0.03 \\
                   &  1    &  --   &  1    &  1    & 4     & 14    &  4    \\
           HD 31867 & -0.21 & -- & +0.13 & -0.06 &-0.13  & -0.07 & -0.05 \\
                   &  0.04 &  -- &  0.05 &  0.14 & 0.04  &  0.01 &  0.03 \\
                   &  1    &  --    &  1    &  1    & 5     & 14    &  6    \\
           HD 33866 & -0.06 & -- & -- & +0.04 &-0.35  & -0.22 & -0.20 \\
                   &  0.06 &  -- &  -- &  0.17 & 0.08  &  0.02 &  0.04 \\
                   &  1    &  --   & --    &  1    & 4     & 15    &  5    \\
           HD 41708 & +0.22 & -- & -- & -0.03 &+0.07  & +0.08 & +0.12 \\
                   &  0.04 &  -- &  -- &  0.16 & 0.05  &  0.01 &  0.03 \\
                   &  1    &  --    &  --    &  1    & 4     & 15    &  6    \\
           HD 42807 & +0.02 & -- & +0.03 & +0.06 &-0.05  & -0.10 & -0.15 \\
                   &  0.04 &  -- &  0.06 &  0.16 & 0.05  &  0.01 &  0.04 \\
                   &  1    &  --    &  1    &  1    & 5     & 15    &  5    \\
           HD 77730 & -0.50 & -0.18 & -0.15 & -0.30 &-0.41  & -0.40 & -0.35 \\
                   &  0.06 &  0.05 &  0.04 &  0.16 & 0.06  &  0.02 &  0.04 \\
                   &  1    &  1    &  1    &  1    & 5     & 16    &  6    \\
          HD 110882 & -0.32 & -- & -- & -0.31 &-0.18  & -0.37 & -0.31 \\
                   &  0.04 &  -- &  -- &  0.14 & 0.05  &  0.02 &  0.04 \\
                   &  1    &  --    &  --    &  1    & 4     &  9    &  6    \\
          HD 110884 & -- & -0.17 & -- & -0.11 &-0.18  & -0.14 & -0.11 \\
                   &  -- &  0.04 &  -- &  0.19 & 0.07  &  0.02 &  0.05 \\
                   &  --    &  1    &  --    &  1    & 3     & 16    &  4    \\
          HD 111513 & +0.23 & +0.03 & +0.21 & +0.08 &-0.03  & +0.04 & +0.12 \\
                   &  0.04 &  0.04 &  0.03 &  0.16 & 0.06  &  0.01 &  0.03 \\
                   &  1    &  1    &  2    &  1    & 4     & 17    &  5    \\
          HD 111540 & +0.00 & +0.17 & +0.28 & +0.24 &+0.09  & +0.13 & +0.14 \\
                   & 0.09  & 0.04  & 0.04  & 0.14  &0.05   & 0.01  & 0.04  \\
                   &  1    &  1    &  2    &  1    & 4     & 15    &  6    \\
          HD 124019 & -0.16 & -0.15 & +0.30 & -0.17 &-0.12  & -0.05 & -0.24 \\
                   &  0.06 &  0.04 &  0.06 &  0.14 & 0.05  &  0.01 &  0.04 \\
                   &  1    &  1    &  1    &  1    & 5     & 13    &  5    \\
          HD 126991 & -0.29 & -0.20 & -0.41 & -0.20 &-0.10  & -0.53 & -0.37 \\
                   &  0.04 &  0.04 &  0.04 &  0.20 & 0.06  &  0.02 &  0.04 \\
                   &  1    &  1    &  1    &  1    & 5     & 16    &  5    \\
          HD 129357 & -0.04 & -0.01 & -- & -0.03 &+0.02  & -0.01 & -0.01 \\
                   &  0.04 &  0.03 &  -- &  0.13 & 0.04  &  0.01 &  0.03 \\
                   &  1    &  1    &  --    &  1    & 5     & 16    &  6    \\
          HD 130948 & -0.08 & -- & +0.11 & -0.10 &-0.28  & -0.11 & -0.14 \\
                   &  0.04 &  -- &  0.06 &  0.16 & 0.08  &  0.02 &  0.04 \\
                   &  1    &  --    &  1    &  1    & 3     & 15    &  3    \\
          HD 135145 & -0.19 & -- & -- & -0.11 &+0.00  & -0.03 & +0.03 \\
                   &  0.04 &  -- &  -- &  0.16 & 0.06  &  0.02 &  0.03 \\
                   &  1    &  --    &  --    &  1    & 4     & 15    &  6    \\

\hline
\end{tabular}
\label{tab:abun}
\end{table*}

\begin{table*}
\contcaption{}
\begin{tabular}{lrrrrrrr} \hline
Star & [Mg/H] & [Al/H] & [Si/H] & [Ca/H] & [Ti/H] & [Fe/H] & [Ni/H]\\ \hline 
          HD 135633 & +0.02 & +0.14 & +0.29 & +0.33 &+0.14  & +0.23 & +0.13 \\
                   &  0.06 &  0.04 &  0.08 &  0.19 & 0.06  &  0.01 &  0.04 \\
                   &  1    &  1    &  1    &  1    & 4     & 15    &  5    \\
          HD 140385 & +0.01 & -0.03 & -0.11 & -0.28 &+0.12  & -0.24 & -0.15 \\
                   &  0.05 &  0.04 &  0.04 &  0.15 & 0.05  &  0.01 &  0.03 \\
                   &  1    &  1    &  2    &  1    & 5     & 16    &  6    \\
          HD 145404 & -0.25 & -0.25 & -- & -0.23 &-0.08  & -0.18 & -0.19 \\
                   &  0.04 &  0.04 &  -- &  0.16 & 0.05  &  0.01 &  0.04 \\
                   &  1    &  1    &  --    &  1    & 5     & 16    &  6    \\
          HD 152264 & -0.11 & +0.00 & +0.03 & +0.16 &+0.17  & +0.07 & +0.10 \\
                   &  0.04 &  0.04 &  0.04 &  0.16 & 0.06  &  0.01 &  0.03 \\
                   &  1    &  1    &  2    &  1    & 4     & 17    &  5    \\
         BD+29 2963 & -0.17 & -0.16 & -0.28 & -0.30 &-0.08  & -0.22 & -0.23 \\
                   &  0.04 &  0.04 &  0.04 &  0.13 & 0.04  &  0.01 &  0.04 \\
                   &  1    &  1    &  1    &  1    & 5     & 17    &  5    \\
          HD 156968 & +0.00 & -0.11 & -- & -0.03 &+0.06  & -0.03 & -0.02 \\
                   &  0.05 &  0.04 &  -- &  0.16 & 0.06  &  0.02 &  0.04 \\
                   &  1    &  1    &  --    &  1    & 5     & 16    &  6    \\
          HD 168874 & +0.00 & +0.03 & +0.05 & +0.00 &+0.04  & -0.01 & +0.07 \\
                   &  0.05 &  0.04 &  0.08 &  0.16 & 0.06  &  0.02 &  0.04 \\
                   &  1    &  1    &  1    &  1    & 5     & 15    &  4    \\
         BD+28 3198 & +0.44 & +0.30 & +0.35 & +0.46 &+0.33  & +0.27 & +0.36 \\
                   &  0.04 &  0.04 &  0.04 &  0.12 & 0.04  &  0.01 &  0.03 \\
                   &  1    &  1    &  2    &  1    & 4     & 18    &  5    \\
          HD 333565 & -0.19 & -0.05 & -0.02 & -0.16 &+0.12  & -0.03 & +0.04 \\
                   &  0.04 &  0.04 &  0.03 &  0.15 & 0.06  &  0.01 &  0.04 \\
                   &  1    &  1    &  1    &  1    & 4     & 13    &  3    \\
          HD 193664 & -0.26 & -- & +0.09 & -0.02 &-0.17  & -0.12 & -0.06 \\
                   &  0.05 &  -- &  0.08 &  0.17 & 0.09  &  0.02 &  0.04 \\
                   &  1    &  --    &  1    &  1    & 3     & 16    &  5    \\
         BD+47 3218 & +0.06 & +0.09 & +0.15 & +0.28 &+0.24  & +0.13 & +0.32 \\
                   &  0.04 &  0.04 &  0.04 &  0.16 & 0.06  &  0.01 &  0.03 \\
                   &  1    &  1    &  2    &  1    & 4     & 15    &  5    \\
          HD 210460 & -0.49 & -- & -- & -0.54 &-0.38  & -0.37 & -0.37 \\
                   &  0.05 &  -- &  -- &  0.17 & 0.06  &  0.02 &  0.04 \\
                   &  1    &  --    &  --    &  1    & 4     & 13    &  5    \\
    TYC 3986-3381-1 & -- & +0.38 & +0.48 & +0.40 &+0.29  & +0.32 & +0.23 \\
                   &  -- &  0.05 &  0.04 &  0.15 & 0.07  &  0.01 &  0.04 \\
                   &  0    &  1    &  1    &  1    & 4     & 14    &  4    \\
          HD 212809 & +0.01 & -- & -- & +0.28 &+0.03  & +0.08 & +0.08 \\
                   &  0.07 &  -- &  -- &  0.15 & 0.05  &  0.01 &  0.04 \\
                   &  1    &  --    &  --    &  1    & 4     & 15    &  5    \\
         BD+28 4515 & +0.02 & -- & +0.38 & +0.24 &-0.10  & +0.07 & +0.10 \\
                   &  0.04 &  -- &  0.05 &  0.13 & 0.05  &  0.01 &  0.03 \\
                   &  1    &  --    &  1    &  1    & 4     & 17    &  6    \\
\hline
\end{tabular}
\end{table*}

\subsection{Super metal rich stars and the [Ref] index}
In our working sample, we included 11 stars considered as SMR ([M/H]~$\geq$~0.16~dex) in 
\cite{lopezval14}. From the present high-resolution analysis, we confirm the SMR status,  by means 
of their iron abundance, for 6 objects, namely BD+60~402, BD+60~600, HD~237200, HD~135633, BD+28~3198, and 
TYC~3986-3381-1, while, for other 3 stars (BD+47~3218, HD~212809, HD~232824), we obtain [Fe/H]
lower than the SMR threshold. However, both BD+47~3218 and HD~212809 have super-solar abundances for all atomic species and some of them are well above the +0.16~dex threshold.
The two remaining cases of SMR stars, HD~228356 and TYC~2655-3677-1, are discussed below. 

The 6 SMR stars are therefore excellent targets to search for giant planet companions. In order to quantify the probability of detecting these planets, we make use of the [Ref] index defined by \cite{gonzalez09}:
\begin{dmath}
{\rm [Ref]}~=~\log\left(24\times10^{7.55+{\rm [Mg/H]}}+28\times10^{7.53+{\rm [Si/H]}}+ 
56\times10^{7.47+{\rm[Fe/H]}}\right)-9.538
\end{dmath}
%\vspace{2cm}

We report the [Ref] index and [Fe/H] in Table~\ref{tab:ppl}, where we also provide the probability 
[$\mathcal{P(\%)}$] of hosting a giant planet, obtained from the probability functions of \cite{gonzalez14} 
and \cite{fischer05}, based solely on chemical composition considerations. BD+60~600 (39$\%$) and BD+28~3198 
(22$\%$) stand out as the best targets for a giant exoplanet search program.

\subsection{[X/Fe] behaviour and comparison with literature data.}
\begin{figure*}
\includegraphics[width=185mm]{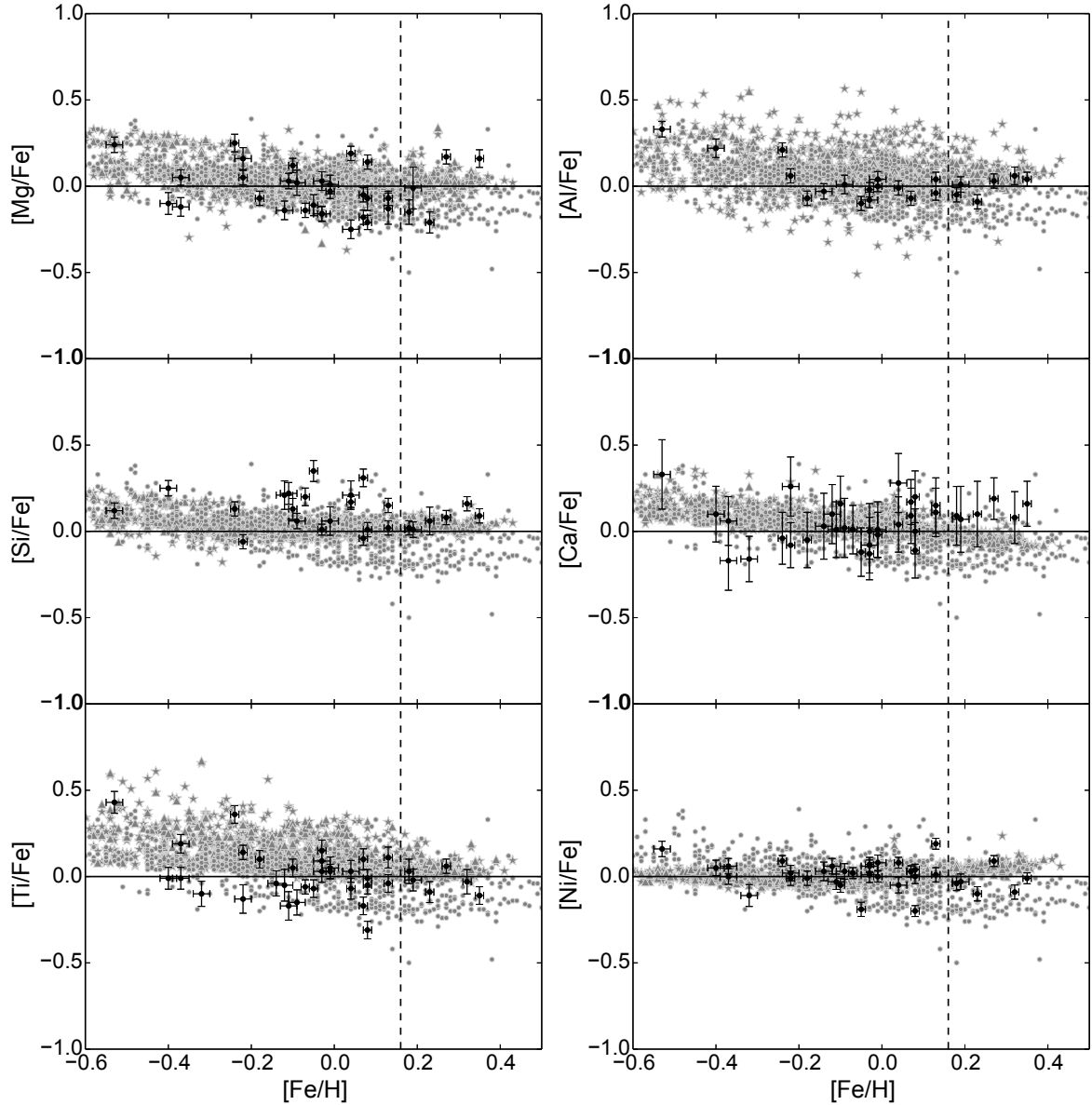} 
\caption{[X/Fe] vs [Fe/H] ratio for our sample (black circles), \citet{neves09}(gray triangles), \citet{adibekyan12} (gray stars), and for \citet{hinkel14} (gray circles). The vertical dashed line indicates the super metallicity threshold.}
\label{fig:abun}
\end{figure*}

In Figure~\ref{fig:abun}, we show the [X/Fe] ratios for the elements included in our analysis.
In order to check for consistency with other abundance studies on objects of the solar neighbourhood, we compare our results with the works of 
\cite{neves09}, \cite{adibekyan12}, and \cite{hinkel14}, which include LTE abundances for FGKM main sequence 
stars, within a distance of 150~pc from the Sun.
We found good agreement with these previous works. 
Our Mg, Si, Ca and Ti ratios present a higher scatter than Al, and Ni, nevertheless, this  pattern is also 
present in the comparison sample.

The errors in the Ca abundance  are, on average, larger than for the other elements and always higher than 
0.10~dex. This anomaly is due to fact that the Ca abundance is very sensitive to the error in surface 
gravity: in fact, we found that $\sigma_{\log{g}}=0.20$~dex produces a difference of 0.08~dex in the Ca 
abundance, while for the other elements the uncertainty in $\log{g}$ does not affect much the overall error.

\begin{figure}
\includegraphics[width=80mm]{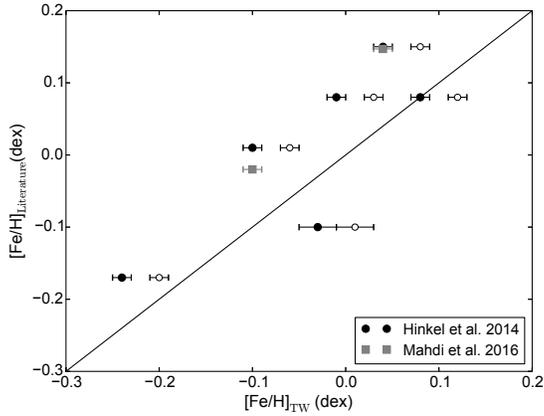}
\caption{Comparison of our iron abundances and those of \citet{hinkel14} (filled circles) and 
\citet{mahdi16} (filled squares). The empty circles represent the transformation of our iron abundances to
the \citet{hinkel14} reference solar abundances.}
\label{fig:comp}
\end{figure}

We found 8 of our stars in the Hypatia catalogue, a compilation of chemical abundances from high-resolution 
spectroscopy \citep{hinkel14}, and 2 objects are also present in the more recent work by \cite{mahdi16}. We found a maximum (minimum) difference of +0.20~dex (-0.02~dex) between our abundances and 
those of \cite{hinkel14}. This discrepancy is as large as the typical dispersion among catalogues included in \cite{hinkel14}. As an example, in Fig.~\ref{fig:comp}, we show the 
comparison of our [Fe/H] values and those of Hypatia for the the stars HD~41708, HD~42807, HD~111513, 
HD~129357, HD~140385, and HD~156968; we also include the iron abundance of \cite{mahdi16} 
for HD~42807 and HD~111513. If we take into account that the solar scale of \cite{lodders09}, used as 
reference by \cite{hinkel14}, has a iron abundance 0.05~dex lower than in \citet{grevesse98}, the agreement 
with our results improves.

\cite{mahdi16} provide the abundance of Si, Ca, Ti, Fe, and Ni for the stars HD~42807 and HD~111513. We found a difference with our results between $+0.08$ and $+0.11$~dex for HD~42807 and in the 
interval $-0.15$ to $+0.07$~dex for HD~111513. Such values, although larger than our errors, can be 
explained by systematic differences, such as different $\log{gf}$ values or different 
atmospheric parameters adopted.

\subsection{Stars with broad line profiles}
\begin{figure}
\centering
\includegraphics[width = 80mm]{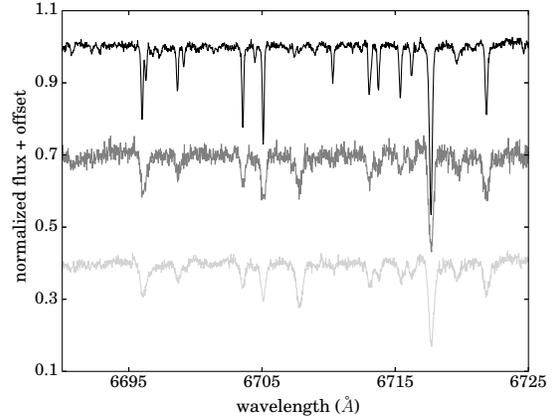}
\caption{The spectra of TYC~2655-3677-1 (gray), and HD~268356 (light gray), compared with the spectrum of Vesta (black).}
\label{fig:rot}
\end{figure}

Two stars, TYC~2655-3677-1 and HD~228356, show line profiles which are significantly broader that the rest 
of the sample (see Fig.~\ref{fig:rot}). This is due to relatively high rotational velocity (with a possible 
significant contribution by macroturbulence). These two objects also have high lithium abundance 
(A(Li)=2.54 for TYC~2655-3677-1 and HD~228653 of A(Li)=2.71; \citealt{rlopezval15}), indicating that 
they are probably young stars.

Their line profiles, however, are broad enough to make very difficult to identify isolated, 
un-blended lines for a correct abundance measurement. We have therefore excluded the two stars from our 
abundance analysis.We measured the FWHM and we computed, using eq. 6 of \cite{strassmeier90}, the projected 
rotation velocity ($v \sin{i}$) for 6 atomic lines in the region around 6710~\AA.

We assumed a macroturbulence velocity of 3~km\,s$^{-1}$ and an instrumental FWHM~=~0.19~\AA\ and we obtained  
$v \sin{i}=$~8.5 and 9.7~km\,s$^{-1}$ for TYC~2655-3677-1 and HD~228653, respectively.

\begin{table}
\caption{Solar abundances from Vesta spectrum and the values of \citet{grevesse98}.}
\label{tab:ref}
\begin{tabular}{ccc}
\hline
Element   & Vesta & G\&S  \\ \hline
Mg & 7.54 $\pm$ 0.02 & 7.58 $\pm$ 0.05  \\
Al & 6.46 $\pm$ 0.02 & 6.47 $\pm$ 0.07\\
Si & 7.52 $\pm$ 0.02 & 7.55 $\pm$ 0.05\\
Ca & 6.38 $\pm$ 0.09 & 6.36 $\pm$ 0.02\\
Ti & 4.98 $\pm$ 0.03 & 5.02 $\pm$ 0.06\\
Fe & 7.49 $\pm$ 0.01 & 7.50 $\pm$ 0.05\\
Ni & 6.23 $\pm$ 0.02 & 6.25 $\pm$ 0.04\\
\hline
\end{tabular}
\end{table}

\begin{table}
\caption{Probability of hosting a giant planet using [Fe/H] and the [Ref] index.}
\begin{tabular}{lcrcr}
\hline
Object & $\rm{[Fe/H]}$ & $\mathcal P(\%)$ & $\rm{[Ref]}$ & $\mathcal P(\%)$ \\
\hline
            HD 5649 & -0.32 &  0 & -- & --\\ 
         BD+60 402 & +0.19 &  7 & +0.19 &  8 \\ 
          HD 16894 & +0.04 &  3 & +0.07 &  3 \\ 
         BD+60 600 & +0.35 & 15 & +0.42 & 39 \\ 
         HD 232824 & -0.09 &  1 & -0.07 &  1 \\ 
         HD 237200 & +0.18 &  6 & +0.15 &  6 \\ 
          HD 26710 & +0.08 &  4 & +0.04 &  2 \\ 
          HD 31867 & -0.07 &  2 & -0.03 &  1 \\ 
          HD 33866 & -0.22 &  1 & -- & --\\ 
          HD 41708 & +0.08 &  4 & -- & --\\ 
          HD 42807 & -0.10 &  1 & -0.03 &  1 \\ 
          HD 77730 & -0.40 &  0 & -0.33 &  0 \\ 
         HD 110882 & -0.37 &  0 & -- & --\\ 
         HD 110884 & -0.14 &  1 & -- & --\\ 
         HD 111513 & +0.04 &  3 & +0.14 &  5 \\ 
         HD 111540 & +0.13 &  5 & +0.15 &  6 \\ 
         HD 124019 & -0.05 &  2 & +0.06 &  3 \\ 
         HD 126991 & -0.53 &  0 & -0.43 &  0 \\ 
         HD 129357 & -0.01 &  2 & +0.09 &  4 \\ 
         HD 130948 & -0.11 &  1 & -0.03 &  1 \\ 
         HD 135145 & -0.03 &  2 & -- & --\\ 
         HD 135633 & +0.23 &  8 & +0.21 &  9 \\ 
         HD 140385 & -0.24 &  0 & -0.13 &  0 \\ 
         HD 145404 & -0.18 &  1 & -- & --\\ 
         HD 152264 & +0.07 &  4 & +0.02 &  2 \\ 
        BD+29 2963 & -0.22 &  1 & -0.22 &  0 \\ 
         HD 156968 & -0.03 &  2 & -- & --\\ 
         HD 168874 & -0.01 &  2 & +0.01 &  2 \\ 
        BD+28 3198 & +0.27 & 10 & +0.34 & 22 \\ 
         HD 333565 & -0.03 &  2 & -0.06 &  1 \\ 
         HD 193664 & -0.12 &  1 & -0.08 &  1 \\ 
        BD+47 3218 & +0.13 &  5 & +0.12 &  5 \\ 
         HD 210460 & -0.37 &  0 & -- & --\\ 
   TYC 3986-3381-1 & +0.32 & 13 & -- & --\\ 
         HD 212809 & +0.08 &  4 & -- & --\\ 
        BD+28 4515 & +0.07 &  4 & +0.17 &  7 \\ 
\hline
\end{tabular}
\label{tab:ppl}
\end{table}

\section*{Acknowledgements}
This research has made use of the SIMBAD data base, operated at CDS, Strasbourg, France. The authors  would like to thank CONACyT for financial support through grants CB-2011-169554 and CB-2015-256961. This research has made use of the SIMBAD data base, operated at CDS, Strasbourg, France.

\label{lastpage}

\end{document}